\documentclass{article}
\usepackage[whileod]{chl-alg} 
\usepackage{amsmath,amssymb}
\newtheorem{theorem}{Theorem}
\newtheorem{example}{Example}

\def\rank{{\rm rank}}
\def\des{{\rm des}}
\begin{document}
\title{A note on the Burrows-Wheeler transformation}
\author{Maxime Crochemore \and Jacques D\'esarm\'enien \and Dominique Perrin}
\maketitle
\begin{abstract}
We relate the Burrows-Wheeler transformation with a result in
combinatorics
on words known as the Gessel-Reutenauer transformation.
\end{abstract}
\section{Introduction}
The Burrows-Wheeler transformation is a popular method used for text
compression \cite{BW}. The rough idea is to encode a text in two passes.
In the first pass, the text $w$ is replaced by a text $T(w)$ of the same length
obtained
as follows: list the cyclic shitfs of $w$ in alphabetic order as the
rows $w_1,w_2,\ldots,w_n$ of an array. Then
$T(w)$
is the last column of the array. In a second
pass, a simple encoding allows to compress $T(w)$, using a simple method
like run-length or move-to-front encoding. Indeed, adjacent rows
will often begin by a long common prefix and $T(w)$ will therefore have
long runs of identical symbols. For example, in a text in english, most rows
beginning with `nd' will end with `a'. We refer to \cite{Manzini} for a complete
presentation
of the algorithm and an analysis of its performances.
It was remarked recently by S. Mantaci, A. Restivo
and M. Sciortino \cite{MRS} that this transformation was related with notions
in combinatorics on words such as Sturmian words. Similar considerations were developped in
\cite{Bannai} in a different context. The results presented here are also
close to the ones of \cite{Duval}.

In this note, we study the transformation from the combinatorial point of
view. We show that the Burrows-Wheeler transformation is a particular case
of a bijection due to I.M. Gessel and C. Reutenauer which allows the enumeration
of permutations by descents and cyclic type (see \cite{Lothaire}).

The paper is organized as follows. In the first section, we describe the
Burrows-Wheeler
transformation. The next section describes the inverse of the
transformation
with some emphasis on the computational aspects. The last section
is devoted to the link with the Gessel-Reutenauer correspondance.
\section{The Burrows-Wheeler transformation}
The principle of the method is very simple. We consider an ordered alphabet $A$.
Let $w=a_1a_2\cdots a_n$ be a word of length $n$ on the
alphabet $A$. The \textit{Parikh vector} of a word $w$ on the alphabet $A$ is the
integer 
vector $v=(n_1,n_2,\ldots,n_k)$ where $n_i$ is the number of occurrences
of the $i$-th letter of $A$ in $w$. We suppose $w$ to be primitive, i.e. that $w$ is not a power
of
another word. Let $w_1,w_2,\ldots,w_n$ be the sequence of conjugates of $w$
in increasing alphabetic order. Let $b_i$ denote the last letter of $w_i$,
for $i=1,\ldots,n$. Then the Burrows-Wheeler transform of $w$ is the word
$T(w)=b_1b_2\cdots b_n$.
\begin{example}{\rm
Let $w=abracadabra$. The list of conjugates of $w$ sorted in alphabetical
order
is represented below.
$$\begin{array}{cccccccccccc}
&1&2&3&4&5&6&7&8&9&10&11\\
1&a&a&b&r&a&c&a&d&a&b&r\\
2&a&b&r&a&a&b&r&a&c&a&d\\
3&a&b&r&a&c&a&d&a&b&r&a\\
4&a&c&a&d&a&b&r&a&a&b&r\\
5&a&d&a&b&r&a&a&b&r&a&c\\
6&b&r&a&a&b&r&a&c&a&d&a\\
7&b&r&a&c&a&d&a&b&r&a&a\\
8&c&a&d&a&b&r&a&a&b&r&a\\
9&d&a&b&r&a&a&b&r&a&c&a\\
10&r&a&a&b&r&a&c&a&d&a&b\\
11&r&a&c&a&d&a&b&r&a&a&b\\
\end{array}$$
The word $T(w)$ is the last column of the array. Thus
$T(w)=rdarcaaaabb$.
}
\end{example}
It is clear that $T(w)$ depends only on the conjugacy class of $w$.
Therefore, in order to study the correspondance $w \mapsto T(w)$, we may suppose that $w$ is a Lyndon word, i.e. that
$w=w_1$. Let $c_i$ denote the first letter of $w_i$. Thus the word
$z=c_1c_2\cdots c_n$ is the nondecreasing rearrangement of $w$ (and of $T(w)$).

Let $\sigma$ be the permutation of the set $\{1,\ldots,n\}$
such that $\sigma(i)=j$ iff $w_j=a_ia_{i+1}\cdots a_{i-1}$.
In other terms, $\sigma(i)$ is the rank in the alphabetic order of the
$i$-th circular shift of the word $w$.
\setcounter{example}{0}
\begin{example}{\rm(continued)

 We have
$$\sigma=\left(\begin{array}{ccccccccccc}
1&2&3&4&5&6&7&8&9&10&11\\
1&3&7&11&4&8&5&9&2&6&10
\end{array}\right)$$
}
\end{example}
By definition, we have for each index $i$ with $1\le i\le
n$
\begin{equation}\label{eq1}
a_i=c_{\sigma(i)}.
\end{equation}
We also have the following
formula expressing $T(w)$ using $\sigma$
\begin{equation}\label{eq2}
b_i=a_{\sigma^{-1}(i)-1}
\end{equation}
Indeed, $b_{\sigma(j)}$ is the last letter of
$w_{\sigma(j)}=a_ja_{j+1}\cdots a_{j-1}$, whence $b_{\sigma(j)}=a_{j-1}$
which is equivalent to the above formula.

Let $\pi=P(w)$ be the permutation defined by $\pi(i)=\sigma(\sigma^{-1}(i)+1)$
where the addition is to be taken $\bmod n$. Actually, $\pi$ is just the
permutation obtained by writing $\sigma$ as a word and interpreting it as
an $n$-cycle. Thus, we have also $\sigma(i)=\pi^{i-1}(1)$ and
\begin{equation}\label{eq2prime}
a_i=c_{\pi^{i-1}(1)}
\end{equation}

\setcounter{example}{0}
\begin{example}{\rm(continued)

 We have, written as a cycle
$$\pi=\left (
\begin{array}{ccccccccccc}
1&3&7&11&4&8&5&9&2&6&10
\end{array}\right )
$$
and as an array
$\pi=\left(\begin{array}{ccccccccccc}
1&2&3&4&5&6&7&8&9&10&11\\
3&6&7&8&9&10&11&5&2&1&4
\end{array}\right)$
}
\end{example}
Substituting in Formula (\ref{eq2}) the value of $a_i$ given by Formula
(\ref{eq1}), we obtain
$b_i=c_{\sigma(\sigma^{-1}(i)-1)}$ which is equivalent to
\begin{equation}\label{eq3}
c_i=b_{\pi(i)}
\end{equation}
Thus the permutation $\pi$ transforms the last column of the array of
conjugates
of $w$ into the first one. Actually, it can be noted that $\pi$ transforms
any column of this array into the following one.

The computation of $T(w)$ from $w$ can be done in linear time. Indeed,
provided
$w$ is chosen as a Lyndon word, the order between the conjugates is the
same
as the order between the corresponding suffixes. The computation of the
permutation $\sigma$ results from the suffix array of $w$ which can be
computed
in linear time \cite{CR} on a fixed alphabet. The corresponding result on
the alphabet of integers is a more recent result. It has been proved
independently by  three groups of researchers, \cite{KSPP}, \cite{KA} and \cite{KS}.
\section{Inverse transformation}
We now show how $w$ can be recovered from $T(w)$. For this, we introduce
the following notation. The rank of $i$ in the word $y=b_1 b_2\cdots b_n$,
denoted $\rank(i,y)$
is the number of occurrences of the letter $b_i$ in $b_1b_2\cdots b_i$.

We observe that for each index $i$, and for the aforementioned
words $y=b_1b_2\cdots b_n$ and $z=c_1c_2\cdots c_n$
\begin{equation}\label{eq4}
\rank(i,z)=\rank(\pi(i),y).
\end{equation}
Indeed, we first note that for two words $u,v$ of the same length and
any letter $a$, one has $au<av\Leftrightarrow ua<va\quad (\Leftrightarrow u<v)$.
Thus for all indices $i,j$
\begin{equation}\label{eq5}
i<j \mbox{ and } c_i=c_j \Rightarrow \pi(i)<\pi(j).
\end{equation}
 Hence, the number of occurrences of
$c_i$ in $c_1c_2\cdots c_i$ is equal to the number of occurrences of $b_{\pi(i)}=c_i$ in
$b_1b_2\cdots b_{\pi(i)}$.

To obtain $w$ from $T(w)=b_1b_2\cdots b_n$, we first compute $z=c_1c_2\cdots c_n$ by
rearranging
the letters $b_i$ in nondecreasing order. Property (\ref{eq4}) shows that
$\pi(i)$
is the index $j$ such that $c_i=b_j$ and $\rank(j,y)=\rank(i,z)$.
This defines the permutation $\pi$, from which $\sigma$ can be
reconstructed. An algorithm  computing $\pi$ from $y=T(w)$
is represented below.

\begin{center}
\begin{algo}{Permutation}{b_1b_2\cdots b_n}
   \SET{c}{\Call{sort}{b_1b_2\cdots b_n}}
   \DOFORI{i}{1}{n}
      \IF{i=1 \mbox{ or } c_{i-1} \neq c_i}
         \SET{j}{0}
      \FI
      \DO
         \SET{j}{j+1}
      \WHILEOD{b_j \neq c_i}
      \SET{\pi(i)}{j}
   \OD
   \RETURN{\pi}
\end{algo}
\end{center}
This algorithm can be optimized to a linear-time algorithm by storing the
first position of each symbol in the word $z$.

Finally $w$ can be recovered from $z=c_1c_2\cdots c_n$  and $\pi$ by Formula (\ref{eq2prime}).
The algorithm allowing to recover $w$  
is represented
below.

\begin{center}
\begin{algo}{Word}{z,\pi}

   \SET{j}{1}
   \SET{a_1}{c_1}
   \DOFORI{i}{2}{n}
      \SET{j}{\pi(j)}
      \SET{a_j}{c_j}
   \OD
   \RETURN{w}
\end{algo}
\end{center}
The computation of $w$ is not possible without the Parikh vector or
equivalently
the word $z$. One can however always compute the word $w$ on the smallest
possible alphabet associated with permutation $\pi$ (this is the
computation
described in \cite{Bannai}).
\section{Descents of permutations}
A descent of a permutation $\pi$ is an index $i$  such that
$\pi(i)>\pi(i+1)$. We denote by $\des(\pi)$ the set of descents of the
permutation $\pi$. It is clear by Property (\ref{eq5}) that if $i$ is a descent of
$P(w)$, then $c_i\neq c_{i+1}$. Thus, the number of descents
of $\pi$ is at most equal to $k-1$ where $k$ is the number
of symbols appearing in the word $w$.
\setcounter{example}{0}
\begin{example}{\rm(continued)
The descents of $\pi$ appear in boldface.
$$\pi=\left(\begin{array}{ccccccccccc}
1&2&3&4&5&6 &\mathbf{7} &\mathbf{8}&\mathbf{9}&10&11\\
3&6&7&8&9&10&11         &5         &2&1 &4
\end{array}\right)$$
Thus $\des(\pi)=\{7,8,9\}$.
}
\end{example}

Let us fix an ordered alphabet $A$ with $k$ elements for the rest
of the paper.  Let $w$ be a word and $v=(n_1,n_2,\ldots,n_k)$ be the Parikh
vector of $w$. We say that $v$ is \textit{positive} if $n_i>0$
for $i=1,2,\ldots,k$. We denote by $\rho(v)$ the set of integers
$\rho(v)=\{n_1,n_1+n_2,\ldots,n_1+\cdots+n_{k-1}\}$. When $v$ is positive,
$\rho(v)$ has $k-1$ elements. Let $\pi=P(w)$ and let $v$ be the Parikh
vector of $w$.
It is clear by Formula \ref{eq5} that we have the inclusion $\des(\pi)\subset \rho(v)$.
\setcounter{example}{0}
\begin{example}{\rm(continued)
The Parikh vector of the word $w=abracadabra$ is $v=(5,2,1,1,2)$ and $\rho(v)=\{5,7,8,9\}$.
}
\end{example}

The following statement results from the  preceding considerations.

\begin{theorem}\label{th1}
For any positive vector $v=(n_1,n_2,\cdots,n_k)$ with $n=n_1+\cdots+n_k$,
 the map $w\mapsto \pi=P(w)$ is one to one from the set
of conjugacy classes of primitive words of length $n$ on $A$ with Parikh
vector $v$
onto the set of cyclic permutations on $\{1,2,\ldots,n\}$
such that $\rho(v)$ contains $\des(\pi)$.
\end{theorem}
This result is actually a particular case of a result stated in
\cite{Lothaire} and essentially due to I. Gessel and C. Reutenauer
\cite{GesselReutenauer}.
The complete result  (\cite{Lothaire}, Theorem 11.6.1 p. 378) establishes
a bijection between words of type $\lambda$ and pairs $(\pi,E)$ where
$\pi$ is a permutation of type $\lambda$ and $E$ is a subset of
$\{1,2,\ldots,n-1\}$
 with at most $k-1$ elements containing $\des(\pi)$. The type of a word $w$ of length
$n$ is the partition of $n$ realized by the length of the factors of its
nonincreasing factorization in Lyndon words. The type of a permutation is the partition
resulting of the length of its cycles. Thus, Theorem \ref{th1} corresponds
to
the case where $w$ is a Lyndon word (i.e. $\lambda$ has only one part) and
$\pi$ is circular. 

We illustrate the general case of an arbitrary word with an example for
the
sake of clarity. For example, the word $w=abaab$ has the nonincreasing factorization
in Lyndon words $w=(ab)(aab)$. Thus $w$ has type $(3,2)$. The corresponding
permutation of type $(3,2)$ is $\pi=(35)(124)$. Actually, the permutation
$\pi$
is obtained as follows. Its cycles correspond to the Lyndon factors of $w$.
The letters are replaced by the rank in the lexicographic order of the
cyclic iterates of the conjugates. In our example, we obtain
$$
\begin{array}{ccccccccc}
1& &a&a&b&a&a&b&\cdots\\
2& &a&b&a&a&b&a&\cdots\\
3& &a&b&a&b&a&b&\cdots\\
4& &b&a&a&b&a&a&\cdots\\
5& &b&a&b&a&b&a&\cdots
\end{array}
$$
We have $\des(\pi)=\{3\}$
which is actually included in $\rho(v)=\{3,5\}$.

We may observe that when the alphabet is binary, i.e. when $k=2$, Theorem~\ref{th1}
 takes a simpler form: the map $w\mapsto P(w)$ is one-to-one
from the set of primitive binary words of length $n$ onto the set
of circular permutations on $\{1,2,\ldots,n\}$ having 
one descent.

In the general case of an arbitrary alphabet, another possible formulation is the following.
Let us say that a word $b_1b_2\cdots b_n$ is \textit{co-Lyndon}
if the permutation $\pi$ built by Algorithm {\sc Permutation} is an $n$-cycle. It is clear
that the map $w\mapsto T(w)$ is one-to-one from the set of Lyndon words
of length $n$ on $A$ onto the set of co-Lyndon words of length $n$ on $A$.

The properties of co-Lyndon words have never been studied and this might
be an interesting direction of research. 
\begin{example}{\rm
The following array shows the correspondance between Lyndon and co-Lyndon
words of length 5 on $\{a,b\}$. The permutation $\pi$ is shown on the right.
$$\begin{array}{ccc}
\mbox{Lyndon}&\mbox{co-Lyndon}&\\
aaaab&baaaa&(12345)\\
aaabb&baaba&(12354)\\
aabab&bbaaa&(13524)\\
aabbb&babba&(12543)\\
ababb&bbbaa&(14253)\\
abbbb&bbbba&(15432)\\
\end{array}$$
}
\end{example}
\bibliographystyle{plain}
\bibliography{BW}

\end{document}